\documentclass[prl,twocolumn,floatfix,showpacs,superscriptaddress]{revtex4}
\usepackage{graphicx}
\usepackage{amsmath}
\begin{document}

\title{Kondo effect of an adsorbed cobalt phthalocyanine (CoPc)
molecule: \\ the role of quantum interference}
\author{G. Chiappe}
\affiliation{Departamento de F\'{\i}sica Aplicada, Unidad Asociada del
Consejo Superior de Investigaciones Cient\'{\i}ficas and
Instituto Universitario de Materiales, Universidad
de Alicante, San Vicente del Raspeig, Alicante 03690, Spain.}
\affiliation{Departamento de F\'{\i}sica J.J. Giambiagi, Facultad de
Ciencias
Exactas, Universidad de Buenos Aires, Ciudad Universitaria,
1428 Buenos Aires, Argentina.}
\author{E. Louis}
\affiliation{Departamento de F\'{\i}sica Aplicada, Unidad Asociada del
Consejo Superior de Investigaciones Cient\'{\i}ficas and
Instituto Universitario de Materiales, Universidad
de Alicante, San Vicente del Raspeig, Alicante 03690, Spain.}
\date{\today}
\begin{abstract}
A recent experimental study  showed that, distorting a  CoPc
molecule adsorbed on a Au(111) surface, a Kondo effect is induced  with a 
temperature higher than 200 K. We examine a model in which
an atom with strong Coulomb repulsion (Co) is surrounded by four 
atoms on a square (molecule lobes),
and two atoms above and below it representing the apex of the STM tip 
and an atom on the gold surface (all with a single, half-filled, atomic orbital). 
The  Hamiltonian is solved exactly
for the isolated cluster, and, after  connecting the leads (STM tip and gold),  
the  conductance is
calculated by  standard techniques.  Quantum interference prevents the existence of
the  Kondo effect when the  orbitals on the square do
not interact (undistorted molecule); the Kondo resonance shows up after 
switching on that interaction.  The  weight of the Kondo resonance
is controlled by the interplay of couplings to  the STM tip and the gold surface, 
and between the molecule lobes.
\end{abstract}
\maketitle

Coupling of localized spins to conduction electrons may lead to a transport anomaly known 
as the Kondo effect \cite{Ko64,He97}. 
This effect, that usually shows up at low temperatures, consists of a sharp peak 
at the Fermi level, whose half-width is known as the Kondo temperature ($T_{K}$),
and a conductance close to one conductance quantum ${\mathcal G}_0=2e^2/h$. 
The Kondo temperature in the case of magnetic impurities
in non-magnetic metals is around 50 K \cite{Ko64}, whereas in  artificial
atoms (quantum dots) is just  a few hundred  mK \cite{Go98,CO98}. 
In a recent experiment \cite{ZL05,Cr05} it has been shown that it is possible to control
the  characteristics, and even the existence, of the Kondo resonance by modifying the chemical
surroundings of a magnetic atom. The experiments were carried out 
on a cobalt phthalocyanine molecule (CoPc) adsorbed on a Au(111) surface. 
Dehydrogenation of this molecule (d-CoPc) by means of voltage pulses from
a Scanning Tunneling Microscope (STM) triggered a Kondo effect
with a rather high Kondo temperature ($T_{K} \approx$ 200 K). This
temperature is even higher than that observed for bare Co adsorbed on a similar surface
\cite{Cr05,MC98}. Besides such a
high $T_{K}$, one of the most remarkable results of \cite{ZL05} is the fact that the undistorted
molecule does not show a Kondo effect, while  it is readily promoted by distorting
the molecule upon dehydrogenation. Topographic images taken by means
of the STM \cite{ZL05} indicated that the CoPc molecule has four almost non-overlapping lobes
symmetrically placed around the Co atom. Dehydrogenation distorts
the molecule and forces those lobes to overlap. In addition it strongly
decreases the distance from the molecule lobes to the gold surface
and increases the Co/gold surface distance in approximately 30\% \cite{ZL05}.

We hereby propose a simple
model that accounts for some of the salient  features of the experiment described above. 
We take a model Hamiltonian on a small atomic arrangement which is
solved exactly, and subsequently connected to semi-infinite chains used
to describe  the STM tip and the gold surface. Fig. 1 depicts 
this atomic arrangement. A central site with a single atomic orbital and a
strong Coulomb repulsion accounts for the Co atom, while the four lobes of the molecule
are described by four atomic orbitals placed on a square whose center is the Co atom.
Two additional orbitals located above and below the Co atom are included to
represent the apex of the STM tip and an atom on the gold surface, respectively. All atomic
orbitals are assumed to be isotropic ($s$-like). The Hamiltonian takes  the form, 
\begin{equation}
{\hat H} = \sum_{i\sigma}\epsilon_i
c^{\dagger}_{i\sigma}c_{i\sigma} +
\sum_{<ij>;\sigma}t_{i,j}c^{\dagger}_{i\sigma}
c_{j\sigma}+
Un_{Co\uparrow}n_{Co\downarrow}
\label{eq:H}
\end{equation}
\noindent where $c^{\dagger}_{i\sigma}$ creates an electron at site
$i$ with $z$-component of the spin $\sigma=\uparrow,\downarrow$, while the 
occupation operator $n_{Co\sigma}$ associated to Co is,
$n_{Co\sigma}=c^{\dagger}_{Co\sigma}c_{Co\sigma}$.
We just consider the half-filling case, i.e., one electron per atomic orbital.

The parameters of this model Hamiltonian are the following. $t_{i,j}$ is the hopping 
between atomic orbitals located on sites $i$ and $j$ (the symbol $<>$ in Eq. (1) indicates
that $i\neq j$), each orbital has an energy $\epsilon_i$, and the local (Hubbard-like) 
Coulomb repulsion on Co is described by $U$.  In particular we use the following
parameters. We take $\epsilon_{Co}=-U/2$ (symmetric case) and the rest of atomic
orbitals lying at zero energy.
The hoppings  incorporated in the model are: $t_{Co,t}$
(Co and the STM tip) $t_{Co,Au}$ (Co and the gold surface) $t_{Co,l}$ 
(Co and the molecule lobes) $t_{l,l}$ (hopping between lobes). Another 
 important parameter of the model is the hopping between the lobes
 and the gold surface $t_{l,Au}$. One  lead is attached to Co and describes the 
STM tip. The other lead (the Au surface) is attached to either Co or the lobes.
\begin{figure}
\vspace{-3.2cm}
\includegraphics[width=3.2in,height=4.5in]{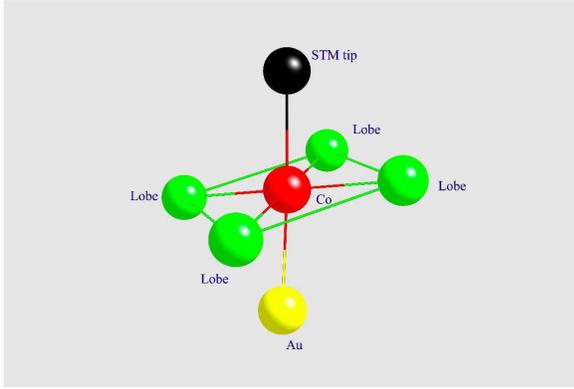}
\vspace{-3.4cm}
\caption{Cluster of atoms utilized to describe the CoPc 
molecule adsorbed on a Au(111) surface. The four atoms on the square account 
for the four  molecule lobes, while the atom at the center represents Co.
The upper (lower) atom accounts for the apex of the tip of the STM microscope 
(the gold surface). To calculate the conductance  semi-infinite chains
were attached to the top atom, and  to either the gold atom or the lobes 
(see text).}
\end{figure}

When the cluster is connected to electrodes, the transmission across the system 
is given by  $T(E)=\frac{2e^2}{h}{\rm Tr}[t^{\dagger}t]$ \cite{La57},
and the conductance is ${\mathcal G}=T(E_F)$, where $E_F$ is
the Fermi level.
In this expression,  matrix $t$ is 
$t= \Gamma_{\rm U}^{1/2}G^{(+)}\Gamma_{\rm L}^{1/2}=
\left[\Gamma_{\rm L}^{1/2}G^{(-)} \Gamma_{\rm U}^{1/2} \right]^{\dagger}$,
where $\Gamma_{\rm U(L)}= i(\Sigma^{(-)}_{\rm U(L)}-\Sigma^{(+)}_{\rm U(L)})$,
$\Sigma^{(\pm)}_{\rm U(L)}$  being the  
self-energies of the upper (U) and lower (L) leads, STM tip and gold surface, 
respectively.  Superscripts (+) and (-)
stand for retarded and advanced.
The Green function  is written as \cite{MW92},
$G^{(\pm)}=\left(\left[G_0^{(\pm)}\right]^{-1}-
\left[\Sigma_U^{(\pm)}+\Sigma_L^{(\pm)}\right]\right)^{-1}$,
where $G_0^{(\pm)}$ is the Green function  associated to the isolated  cluster, 
which is obtained by  exact diagonalization  \cite{FC99,CV03}.
The electrodes  are described by means of semi-infinite chains. 
This method is {\it exact} only as far as the calculation of the Green function 
of the isolated cluster is concerned, and does not account for correlation 
effects that extend beyond its bounds. The method has already been applied to a 
variety of transport problems \cite{FC99,CV03,AD05,BM04,MB06} 
including transport through hydrogenated Pt nanocontacts \cite{CL05}.

In carrying out calculations  we have taken the hopping within
the semi-infinite chains (leads)  to be 1 eV, which is slightly larger than the hopping
between $s$-orbitals in gold \cite{Pa76}. In addition we take  U=8 eV, not far from the value 
6-7 eV recently estimated for the CuPc molecule \cite{Flpc}. The rest of the 
model parameters 
have been varied aiming to identify their role in the behavior of this system.
All calculations were done at zero temperature.
\begin{figure}
\includegraphics[width=3.2in,height=3.8in]{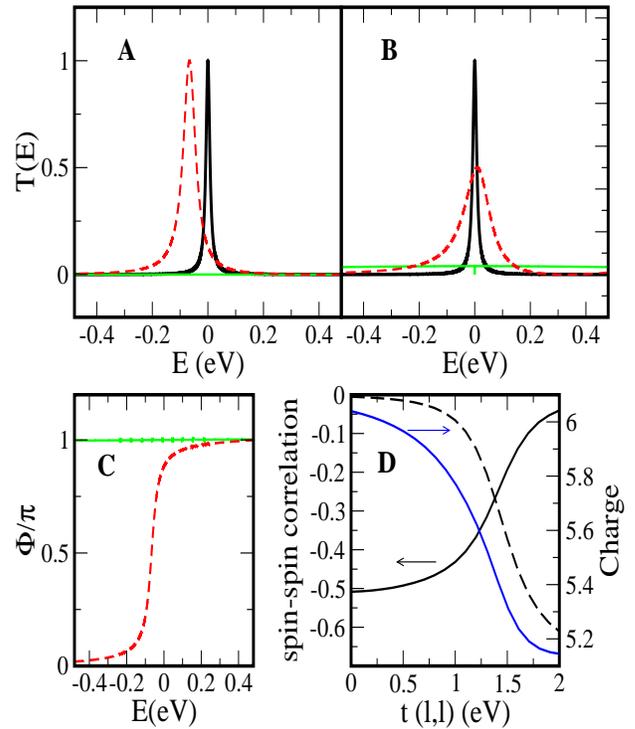}
\caption{Transmission {\it T} (in units of the conductance quantum) versus 
the energy {\it E} (in eV) referred to the Fermi energy, 
and phase  difference $\Phi$ between the direct path from the gold surface 
to the STM tip
going through the Co atom, and the path that passes through the molecule lobes (see text). 
A: $t_{l,Au}=0$, $t_{Co,t}=t_{Co,Au}$=0.25 eV and $t_{Co,l}$=1.25 eV;  
$t_{l,l}=0$  (green),  and $t_{l,l}$= 2 eV  (red).  B:  $t_{Co,Au}=0$,  
$t_{Co,t}$=0.25 eV, $t_{l,Au}$= 1 eV,  and, $t_{Co,l}$=1.25 eV; 
$t_{l,l}=0$ (green),  and $t_{l,l}$= 2 eV (red). 
The standard Kondo resonance, obtained with $t_{Co,l}$= 0 and 
$t_{Co,t}=t_{Co,Au}$=0.25 eV (black),  is plotted in both (A and B).
C: phase difference calculated for the parameters used in A.
D: spin-spin correlation for Co/lobes (continuous line) and Co/STM tip (broken
line)  and total charge on the lobes plus Co (continuous blue line) 
versus $t_{l,l}$, for the parameters used in A.}
\end{figure} 

The results depicted in Fig. 2 are addressed to identify the origin of the
emergence of the Kondo resonance upon distortion of the molecule 
\cite{note1,CN04,LV03}.
The Figure shows results for either the Co atom (Fig. 2A) or the lobes (Fig. 2B) connected 
to the gold surface.  In addition we take either non-interacting lobes 
(green curves) or a finite hopping between lobes (red  curves). 
For comparison we also show the results for the standard Kondo effect in which 
the Co is decoupled from the 
molecule lobes (black curves). The most appealing result is that
{\it when the molecule lobes do not interact no Kondo effect shows up} 
\cite{note2}. The origin of this remarkable result is likely related to quantum 
interference \cite{He97,AD05,CL05,TK05,KC01,BS00,CB03}.  
To illustrate this assessment we have calculated the phase 
difference between the direct path from the gold surface to the STM tip
going through the Co atom, and the path that passes through the molecule lobes.
This phase difference can be 
derived from the  following  element of the Green function,
\begin{eqnarray}
G^{(+)}(Au,t)&=& g^{(+)}(Au,t)+ \nonumber \\
&&4g^{(+)}(Au,Co)\Sigma(Co,l)G^{(+)}(l,t)
\end{eqnarray}
\noindent where $\Sigma(Co,l)$ is a  many-body self-energy
that accounts for the lobes/Co coupling, and lower case "{\it g}" are the Green 
functions in the case that  lobes and  Co are 
decoupled,  (note that $\Sigma(Co,l)\propto t(Co,l)$) \cite{note3}. 
The results shown in Fig. 2C are just the phase difference
between the two terms in the r.h.s of Eq. (2).  When there is no hopping between 
lobes, 
the phase difference is $\pi$ indicating that the two terms  may totally cancel each other
as actually occurs \cite{note4}. An alternative way to look at this
issue is to calculate the local density of states (LDOS) on the Co atom that is obtained
from the  diagonal element of the Green function,
\begin{eqnarray}
G^{(+)}(Co,Co)&=&(\omega-\epsilon_{l})g^{(+)}(Co,Co)\times \nonumber \\
&&\left[(\omega-\epsilon_{l})\Sigma^{2}(Co,l)g^{(+)}(Co,Co)\right]^{-1}
\end{eqnarray}
\noindent where  $\omega$ is the energy referred
to the Fermi energy. It is readily seen that when the lobe orbitals lie at the Fermi
energy ($\epsilon_{l}=0$) the Green function $G^{(+)}(Co,Co)$ vanishes
at that energy and, thus, the LDOS at the Co atom  \cite{note5,GV87}. 
A null density of states
at the Fermi energy on the strongly correlated Co atom implies that no Kondo resonance
should show up, in accordance with the phase analysis. 
Full cancellation of the two terms in the r.h.s.  Eq. (2) is removed when the lobes orbitals 
do not lie exactly at $E_{F}$, a result that can  be derived from Eq. (3).
When coupling between lobes is switched on, the  phase difference
is no longer $\pi$ (Fig. 2C)  and the Kondo resonance shows up (Fig. 2A). 
This results from
the fact that switching on that coupling opens new paths for the electrons
to go from the lobes to Co that contribute  to the phase difference. 
Besides, the peak width becomes significantly widened with respect to the standard
Kondo effect (black curve in Fig. 2A) despite of the fact that the molecules lobes
are not connected to the gold surface.  Fig. 2D shows the spin-spin correlation
for  Co/lobes and Co/STM tip (a similar result is obtained for Co/Au).
Remarkably, switching on the lobe/lobe coupling
shifts the antiferromagnetic correlation from the Co/lobes to the Co/STM tip 
(in the undistorted molecule
the spin on the Co orbital is screened by the spin on the lobes).
In addition, six electrons are localized in
the undistorted molecule (lobes plus Co) a number that is reduced down to five
when the lobe/lobe coupling is switched on (see Fig. 2). 
These results are  consistent with the
existence (absence) of a magnetic moment on the distorted (undistorted) 
molecule,  as derived from the {\it ab initio} calculations
reported in \cite{ZL05}. 
We believe that 
the  mechanism hereby put forward 
for switching on and off the Kondo resonance may apply to a variety of situations. 
\begin{figure}
\includegraphics[width=3.2in,height=3.5in]{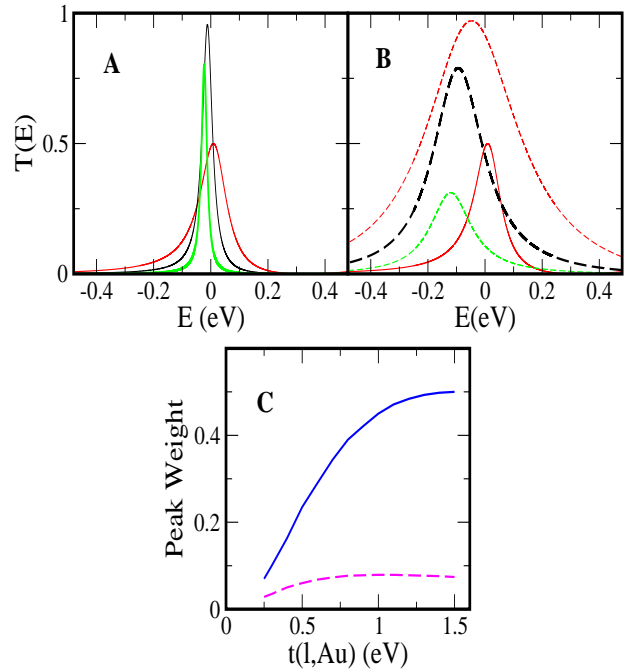}
\caption{Transmission {\it T} (in units of the conductance quantum) versus 
energy {\it E} (in eV) for the Co uncoupled from the gold surface, i.e., $t_{Co,Au}=0$,
and  $t_{Co,l}=1.25$ eV, and $t_{l,l} = 2$ eV. 
A: $t_{Co,t}=0.25$ eV (continuous lines); 
$t_{l,Au}=0.25$ eV (green)  $t_{l,Au}=0.5$ eV (red) and $t_{l,Au}=1$ eV (black).
B: $t_{Co,t}=0.5$ eV (broken lines); 
$t_{l,Au}=0.25$ eV (green)  $t_{l,Au}=0.5$ eV (red) and $t_{l,Au}=1$ eV (black).
One of the curves in A has also been plotted in B for the sake of comparison.
All energies referred to the Fermi energy. C: Peak weight of the Kondo resonance
(in ${\mathcal G}_0$ eV,  ${\mathcal G}_0$ being the quantum of conductance) 
versus the molecule lobes/gold surface hopping, for the two values of $t_{Co,t}$
of A and B (continuous line 0.25 eV and broken line 0.5 eV).}
\end{figure}

Cutting out the  bond between 
Co and the gold surface, and switching on those from lobes to gold, does not
qualitatively change these results. Again, as shown in Fig. 2B, in the absence of lobe-lobe
coupling, no Kondo effect shows up. We note that even though
the lobes/Au surface hopping in Fig. 2B is much higher than the Co/Au surface hopping 
used to obtain the results of Fig. 2A, the width of the resonance is similar and the peak
height considerably smaller (compare the red curves in those two Figures). 
These results suggest that providing more ways to hybridize the 
atomic orbitals on the magnetic ion to the continuum sates (as may be in principle occur due
to coupling of lobes to the gold surface) may not  inevitably be beneficial as 
far as the Kondo effect is concerned. 

The effects of coupling to the leads on the width of the Kondo resonance are illustrated
in Fig. 3. All results correspond to the set of parameters we use to describe the
distorted molecule: finite coupling between the molecule lobes and no hopping between the 
Co atomic orbital and the gold surface. In addition we note that,
in view of the experimental information reported in \cite{ZL05}, it seems
reasonable that the lobe/Au hopping be greater than that related to the Co/STM tip coupling.
Both Fig. 3A and Fig. 3B, show the results for fixed coupling of Co to 
the STM tip and a variable hopping between the lobes and the gold surface. For
the lower value of the Co/STM tip hopping (Fig. 3A),  it is noted
that while the the width of the Kondo resonance follows the expected qualitative 
trend,  it depends only weakly on coupling.
The effect of quantum interference is   demonstrated by the presence of a
non-unitary Kondo effect characterized by a conductance 
smaller than 1 (see also Fig. 2).
A far more important effect on the width of the Kondo peak is obtained when the 
coupling between the Co atomic orbital and the orbital at the apex of the STM 
tip is increased (see Fig. 3B). Now the same increase in the hopping parameter 
produces a dramatic broadening of the Kondo resonance. 
Fig. 3C illustrates
how the weight of the Kondo resonance evolves with the lobe/Au coupling, for
the two values of the Co/STM tip hopping $t(Co,t)$ of Figs. 3A and 3B. 
For the smaller value of $t(Co,t)$ (=0.25 eV), the weight increases only 
slightly with the lobe/Au coupling, saturating around 0.07 ${\mathcal G}_0$ eV for
$t(l,Au) \approx $ 0.7 eV. Instead, for $t(Co,t)$ = 0.5 eV 
the peak weight increases steeply,  saturating for $t(l,Au) > $ 1 eV at 
around 0.5 ${\mathcal G}_0$ eV. These results indicate that both couplings
are equally important, and that, in order to increase the weight of the Kondo 
resonance, the STM tip has to get as closer as possible to the Co atom.
 We note that the results of Fig. 3 that more closely resemble
the experimental data \cite{ZL05}, as far as the peak width is concerned,  
are those corresponding to the lower value of the $t(Co,t)$ hopping 
(compare Fig. 3A of this work and Fig. 2A of Ref. \cite{ZL05}). 
 
An interesting feature of our results is that upon switching on the hopping 
between the molecule lobes,
the electron-hole symmetry is broken and the Kondo resonance is no longer
peaked at the Fermi level. The peak is displaced either upwards or downwards
depending on  the sign of that hopping. 
As we have  assumed that all orbitals are s-like, all hoppings are 
positive.  Interestingly enough this shifts the Kondo peak below the Fermi energy, 
in agreement with the experiments \cite{ZL05}. 
We finally note that if the calculations
of Fig. 3A are done for one lobe decoupled from the other three (a situation
that may represent the case of Fig. 1E of Ref. \cite{ZL05}), a transmission 
smaller than 0.1 ${\mathcal G}_0$ is obtained. It would be interesting to 
check experimentally this prediction.

\acknowledgments
We are grateful to F. Flores, E. Tosatti and J.A. Verg\'es for useful 
discussions and comments. 
Financial support by the spanish MCYT (grants FIS200402356, MAT2005-07369-C03-01 
and NAN2004-09183-C10-08), the Universidad de Alicante, the Generalitat Valenciana
(grant GRUPOS03/092 and grant GV05/152), the Universidad de Buenos
Aires (grant UBACYT x115) and the argentinian CONICET is gratefully acknowledged. 
GC is thankful to the spanish "Ministerio de Educaci\'on y Ciencia" for a Ram\'on
y Cajal grant.


\begin{thebibliography}{99}
\bibitem{note1}
The magnitude actually  plotted in the experimental study of Ref. {\protect \cite{ZL05}} 
is the derivative 
of the  current with respect to the bias potential versus the bias potential. 
However, for small bias potential (as in the present case), that magnitude is 
very similar to the transmission versus energy shown in this work 
(see Refs. {\protect \cite{CN04,LV03}})
\bibitem{note2}
Obviously, our model cannot account for the broad peak observed in the experiment
at around 0.15 eV below the Fermi level, that was related by the authors of 
{\protect \cite{ZL05}} to $d_{z^{2}}$ orbital mediated tunneling.
\bibitem{note3}
Note that the conductance is proportional to the square
of the modulus of the Green function given in Eq. (2).
\bibitem{note4}
In resonance, the two terms in Eq. (2) have the same module and what matters  in
fact is the phase difference.
\bibitem{note5}
A similar effect was reported in Ref. {\protect \cite{GV87}} 
when investigating the effect of linear
side branches on the transmission coefficient in a one-dimensional chain.
\end{thebibliography}
\end{document}